\documentclass[%
 reprint,
superscriptaddress,
 amsmath,amssymb,
 aps,
]{revtex4-1}

\usepackage{graphicx}
\usepackage{dcolumn}
\usepackage{bm}
\usepackage{epstopdf}
\usepackage{xcolor}
\usepackage{epsfig}
\usepackage{float}
\usepackage{stmaryrd}
\usepackage{wasysym}

\begin{document}

\title{Residual stress distributions in athermally deformed amorphous solids from atomistic simulations}
\author{C\'eline Ruscher}
\affiliation{Department of Physics and Astronomy and Quantum Matter Institute, University of British Columbia, Vancouver BC V6T 1Z1, Canada}
\affiliation{Institut Charles Sadron, 23 rue du Loess, F-67034 Strasbourg, France}
\author{J\"org Rottler}
\affiliation{Department of Physics and Astronomy and Quantum Matter Institute, University of British Columbia, Vancouver BC V6T 1Z1, Canada}

\begin{abstract}The distribution of local residual stresses (threshold to instability) that controls the statistical properties of plastic flow in athermal  amorphous solids is examined with an atomistic simulation technique. For quiescent configurations, the distribution has a pseudogap (power-law) form with an exponent that agrees well with global yielding statistics. As soon as deformation sets in, the pseudogap region gives way to a system size dependent plateau at small residual stresses that can be understood from the statistics of local residual stress {\em differences} between plastic events. Results further suggest that the local yield stress in amorphous solids changes even if the given region does not participate in plastic activity. 
\end{abstract}


\renewcommand*\rmdefault{bch}\normalfont\upshape
\rmfamily
\section*{}
\vspace{-1cm}
\maketitle


\footnotetext{\textit{$^{a}$~Institut Charles Sadron, 23 rue du Loess, F-67034 Strasbourg, France,  E-mail: celine.ruscher@ics-cnrs.unistra.fr}}
\footnotetext{\textit{$^{b}$~Department of Physics and Astronomy and Quantum Matter Institute, University of British Columbia, Vancouver BC V6T 1Z1, Canada, E-mail: jrottler@physics.ubc.ca }}




\section{Introduction}
Linking atomic scale observables quantitatively to parameters in effective, coarse-grained descriptions is an important challenge in materials physics. Amorphous materials might appear homogeneous beyond local packing effects, but exhibit structural heterogeneity at the nanoscale \cite{Tsamados2009}. The shear modulus, for instance, is not uniform everywhere but exhibits Gaussian fluctuations on the scale of $\sim 10$ particle diameters that have their origin in the strong nonaffine atomic displacement field of disordered packings already in the elastic regime \cite{Wittmer_2002}.

While local linear properties such as moduli can be probed without an external perturbation through the use of fluctuation formulas \cite{vanWorkum2004}, no such options exists for nonlinear, plastic properties. Ideally one would like to induce plastic yielding in a small local region of interest while eliminating plasticity everywhere else. The only computational approach that has been proposed to date to accomplish this task consists in the (rather drastic) step of enforcing fully affine deformation everywhere but in the region of interest, where particles can move freely \cite{Sollich}. As a result, the nonaffine displacement field (with correlations extending over many particle diameters) is truncated at the boundary between the probe region and the "frozen matrix" (FM). At the linear response level, it is well documented that this truncation shifts the average shear modulus to higher values and reduces the variance of the Gaussian fluctuations in comparison to a fully unconstrained approach \cite{MizunoBarrat2013}. However, the method still faithfully detects about the distribution of soft vs.~hard regions in the amorphous solid. 

For this reason, the FM technique has recently been applied to study the distribution of local yield stresses $\sigma_Y$ in amorphous solids \cite{Francesco2015,Shang_2018}.  Patinet et al. have constructed local yield stress maps of two-dimensional amorphous mixtures after a quench from the liquid phase \cite{Patinet2016,BarbotPatinet2018,Patinet2020}. They found that when these solids are sheared quasistatically, the first plastic events are indeed occurring in regions that exhibit a particularly low local yield stress in the given shear direction. Despite the artefacts potentially induced by the frozen constraint, key information is thus revealed about the distribution of energy barriers that control shear rearrangements. This situates the FM method well within the broader effort to predict the location of irreversible (dynamical) rearrangements from structural (static) features of amorphous packings \cite{manning2012,mosayebi2014,Ding2014,ROYALL20151}. 

Motivated by this success, we seek to explore in this contribution the potential of the FM method to reveal robust statistical information of local mechanical observables in quiescent and flowing amorphous solids. Our focus here is not the yield stress per se, but instead the residual stress $x=\sigma_Y-\sigma_0$, which is the difference between the local yield stress $\sigma_Y$ and the local stress $\sigma_0$ and is a measure of how far a local region is from instability. The distribution $P(x)$ plays a central role in contemporary descriptions of the statistical properties of the yielding transition, because its behavior at small arguments controls the statistics of macroscopic slip events (avalanches) under athermal quasistatic deformation. The distribution is assumed to be scale-free (i.e has a power-law form) and to vanish at zero,  i.e. $P(x)\sim x^{\theta}$ as $x \rightarrow 0$ and the pseudogap exponent $\theta$ enters various scaling relations linking critical exponents of the yielding transition.

In the following, we first present results for $P(x)$ from the FM method applied to the quenched state of a 2D model amorphous solid. We show that a pseudogap form is indeed obtained with an exponent that agrees with one inferred from macroscopic deformation. We then extend the analysis to the transient and steady state of quasistatic deformation, where the distribution $P(x)$ develops a system size dependent plateau in the small $x$ region. By studying the distribution of stress increments $P(\Delta x)$ between slip events, where $\Delta x = \Delta \sigma_Y - \Delta \sigma_0$, we show that this plateau can be attributed to the discrete increments of the underlying mechanical noise. We argue that the FM method reveals correct generic trends in the residual stress distribution $P(x)$, but a quantitative prediction of the pseudogap exponent under deformation requires larger system sizes that are currently computationally not accessible due to combined effects of frozen boundary and finite size artefacts. 

\section{Simulation methods}

\subsection{System}

We consider a 2D Lennard-Jones (LJ) glass-forming binary mixture which has been introduced by Lan\c con {\it{et al.}} \cite{Lancon1988} originally to investigate the properties of 2D quasicrystals. Following refs.~\cite{FalkLanger1998, BarbotPatinet2018}, the $N_L$ large and $N_S$ small particles interact through the potential:
\begin{align}
U_{ab}(r) = \left \{
 \begin{array}{ll}
 \displaystyle
	4 \varepsilon_{ab} \left[ \left(\frac{\sigma_{ab}}{r} \right)^{12} - \left(\frac{\sigma_{ab}}{r} \right)^6 \right] + U_S,  \hspace{0.2cm} \forall \hspace{0.2cm} r \le r_{in} \\
	 \displaystyle \sum_{k=0}^4 C_k r^k \hspace{0.2cm} \forall \hspace{0.2cm},  r_{in} < r \le r_{cut} \\
 \displaystyle	0, \hspace{0.2cm} \forall \hspace{0.2cm} r > r_{cut}
	\end{array}
	\right .
\end{align}
where $\{a,b \}=\{L,S \}$ and $r$ is the distance between two particles.  The potential is shifted at the cutoff distance $r_{cut}=2.5 \sigma_{LS}$ and smoothed for $r_{in} < r \le r_{cut} $ where $r_{in}=2.0 \sigma_{LS}$ in order to ensure that $U_{ab}(r)$ is twice differentiable. 
The shift in energy $U_S$ and the coefficients $C_k$ are:
\begin{align}
U_S &= C_0 -  4 \varepsilon_{ab} \left[ \left(\frac{\sigma_{ab}}{r_{in}} \right)^{12} - \left(\frac{\sigma_{ab}}{r_{in}} \right)^6 \right] \\ \nonumber
C_0 &= -(r_{cut}-r_{in})[3C_1 + C_2(r_{cut}-r_{in})]/6 \\ \nonumber
C_1 &= 24 \varepsilon_{ab} \sigma^6_{ab} (r^6_{in} - 2\sigma^6)/r^{13}_{in} \\ \nonumber
C_2 &= -12 \varepsilon_{ab} \sigma^6_{ab} (7r^6_{in} - 26\sigma^6)/r^{14}_{in} \\ \nonumber	
C_3 &= -[3 C_1 + 4C_2(r_{cut}-r_{in}) ]/[3(r_{cut}-r_{in})^2]   \\ \nonumber	
C_4 &= [C_1 + C_2(r_{cut}-r_{in})]/[3(r_{cut}-r_{in})^3] 
\end{align}

The different LJ parameters are $\sigma_{LL}=2 \sin(\pi/5)$, $\sigma_{LS}=1$, $\sigma_{SS}=2 \sin (\pi/10)$, $\varepsilon_{LL}=\varepsilon_{SS}=0.5$, $\varepsilon_{LS}=1$, and all masses are set to $m=1$. The ratio between large and small particles is chosen such as $N_L/ N_S = (1+ \sqrt{5})/4$, and we work at constant density $N/V=1.0206$. In what follows, the length, mass, energy and time units are expressed in term of $\sigma_{LS}$, $m$, $\varepsilon_{LS}$ and $\sigma_{LS}\sqrt{m/\varepsilon_{LS}}$ respectively. For this system, the glass transition temperature is $T_g=0.325 \varepsilon_{LS}/k_B$ where $k_B$ is Boltzmann's constant.

All simulations have been carried out using the LAMMPS software \cite{lammps}. We consider a 2D triclinic simulation box of size $L$ under periodic boundary conditions where $L \in [53, 300]$ in order to probe system size scaling. To generate the different glass configurations, we first equilibrate systems in the liquid phase at $T=2\,T_g$ using the Langevin thermostat with a damping parameter $T_{damp}=1.0$. The timestep is chosen as $\delta t=0.005$. After equilibration, the different configurations are cooled down to $T=0$ at $dT/dt=2 \cdot 10^{-3}$. Finally an energy minimization is perfomed to ensure that the system is in its local minimum.

\begin{figure}[t]
\centering
\includegraphics[scale=0.10]{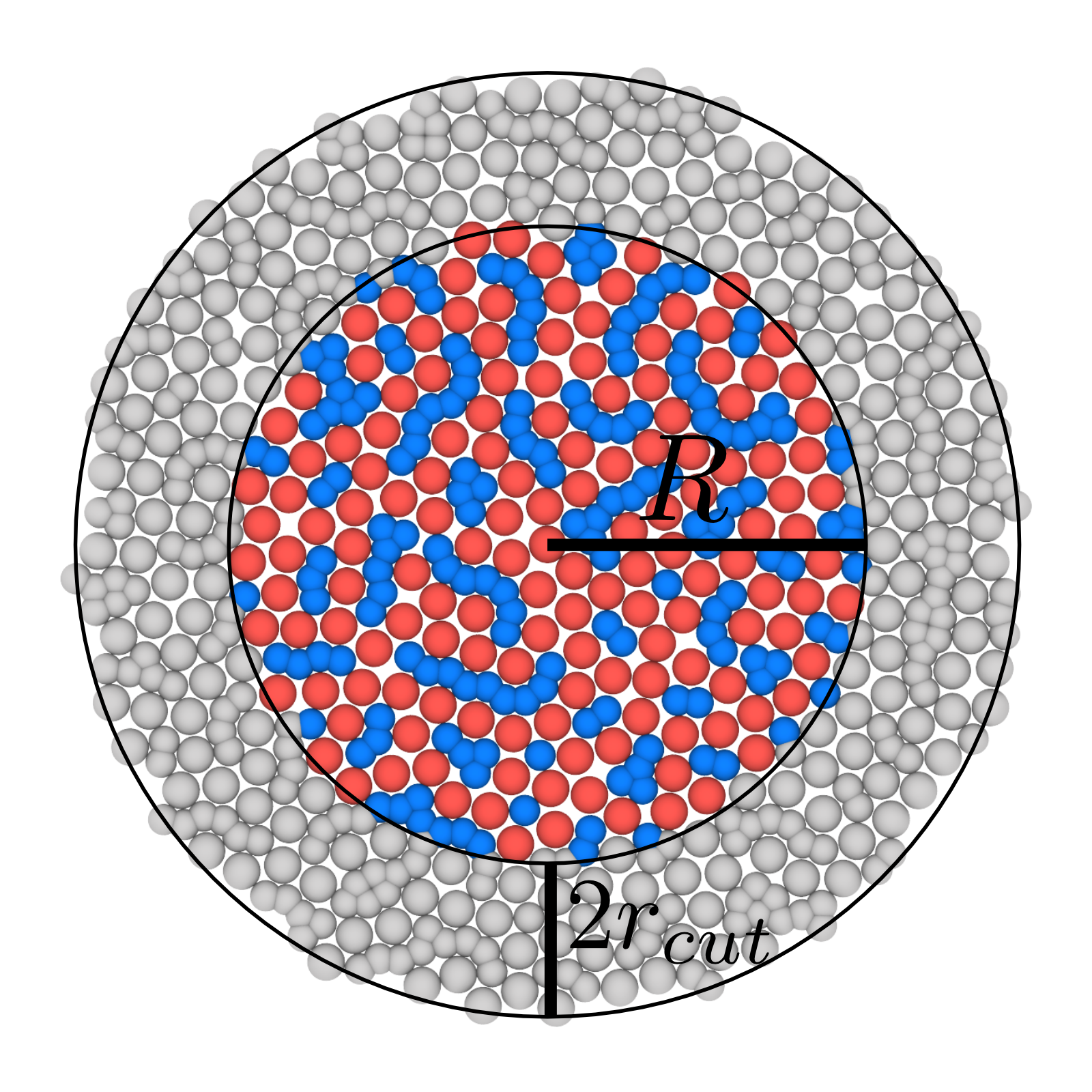}
\includegraphics[scale=0.13]{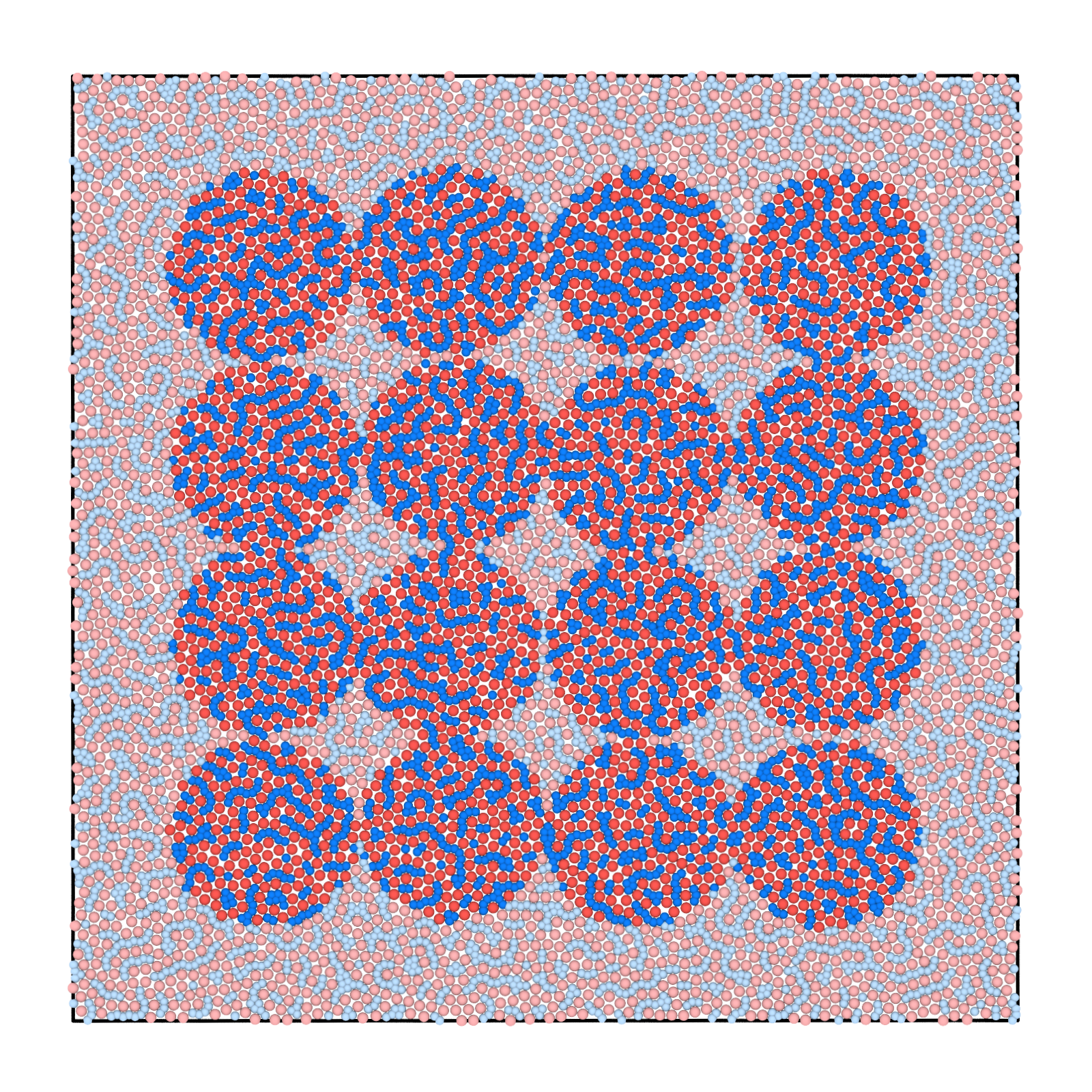}
\caption{Top: Illustration of the frozen matrix method. A circular region of size $R$ is selected from a global configuration and a shell of frozen particles of size $2r_{cut}$ is kept in addition. Bottom: Example of a quenched (applied strain $\gamma=0)$ global configuration of size $L=100$. The circular regions onto which the frozen matrix is applied, are represented with darker colors.The regions do not interpenetrate to ensure their independence during FM analysis.}
\label{fig:FM_illustration}
\end{figure}

\subsection{Athermal quasistatic shear}

Subsequent to cooling, the systems are deformed following the athermal quasistatic shear (AQS) protocol \cite{}. Simple shear is applied in the following way: an affine deformation is first performed by tilting the simulation box in the $x$ direction by an amount $\delta \gamma_{xy} L$, where $\delta \gamma_{xy}= 5 \cdot 10^{-5}$ is the strain increment, and then remapping the position of the particle inside the deformed box. In a second step, we allow the system to relax through an energy minimization using the conjugate gradient method. 

\subsection{Frozen matrix method}

To probe the local properties of our samples, we consider circular regions of size $R$ embedded into a frozen shell of size 2$r_{cut}$ as illustrated in Figure \ref{fig:FM_illustration} (top). During an AQS step, only the inner region can relax implying that if a plastic event occurs, it is necessarily located inside the circular region. In the following, we refer to deformations of the whole periodic simulation box as \emph{global} and to deformations of the circular regions as \emph{local}.

In practice, starting from a global configuration obtained at a given applied strain $\gamma$, we select regions, as shown in Figure \ref{fig:FM_illustration} (Bottom), that we deform under simple shear. By monitoring for each site region the evolution of stress as a function of the applied strain, we can access $\sigma_Y$ and $\gamma_Y$, the local yield stress and yield strain, respectively. Knowing the initial local stress $\sigma_0$ of the circular region, we can determine the residual stresses $x$ associated with each local region. 

\subsection{Detection of plastic events}

To detect plastic events, we revisit an energy-based criterion introduced in  ref.~\cite{Lerner2009} that suits perfectly the AQS protocol. The observable  $\kappa=(U_{\text{aff}}-U_{0})/(N \delta \gamma^2)$ measures the mismatch between the energy associated with the affine displacement $U_{\text{aff}}$ and the inherent structure $U_{0}$. As we show in the Appendix, it is possible to determine analytically an upper bound to $\kappa \le G_B/(2\rho)$ below which system behaves only elastically. $G_B$ is the Born shear modulus. As it is associated to affine deformation, $G_B$ has the same value for both global and local deformations and it is therefore a convenient criterion that can be applied to determine plastic activity in both cases. We numerically found that $G_B \approx  26$, therefore in this work, we consider events that have $\kappa \ge 30 $ as plastic.

\section{Distribution of residual stresses}
In order to assess the ability of the FM method to accurately obtain the distribution $P(x)$, we focus first on the quenched state. We prepare $N_g=5 \cdot 10^{4}$ quenched global configurations of size $L=100$, and from these $N_g$ configurations, we select $N_{\ell} \ge 2 \cdot 10^5$ independent local sites onto which we apply the FM method by deforming these local sites up to the first plastic event. In order to probe the influence of the size $R$ of the local region, we consider $R \in [5.0; 20.0]$. 

Additionally, we deform all $N_g$ configurations globally without any constraints and record the global residual stress $x_{min}$ of the first time failure is observed as well as the associated global yield strain $\gamma_{Y}$. In the AQS protocol, the first failure event represents by definition the weakest site in the system.  According to extreme value statistics, if the underlying distribution of the observable $y=\{x, \gamma_{Y}\}$ is power-law distributed at small arguments, $P(y) \sim y^{\theta}$, then the distribution of independent minimal values $y_{min}$ sampled from  $P(y)$ is expected to follow a Weibull distribution: \cite{Karmakar2010Rapid}
\begin{align}
P(y_{min}) = \frac{1+\theta}{\langle y_{min} \rangle} \left( \frac{y_{min}}{\langle y_{min} \rangle} \right)^{\theta} \exp \left[- \left( \frac{y_{min}}{\langle y_{min} \rangle} \right)^{1+\theta}\right]
\end{align}

\begin{figure}[t]
\centering
\includegraphics[scale=0.12]{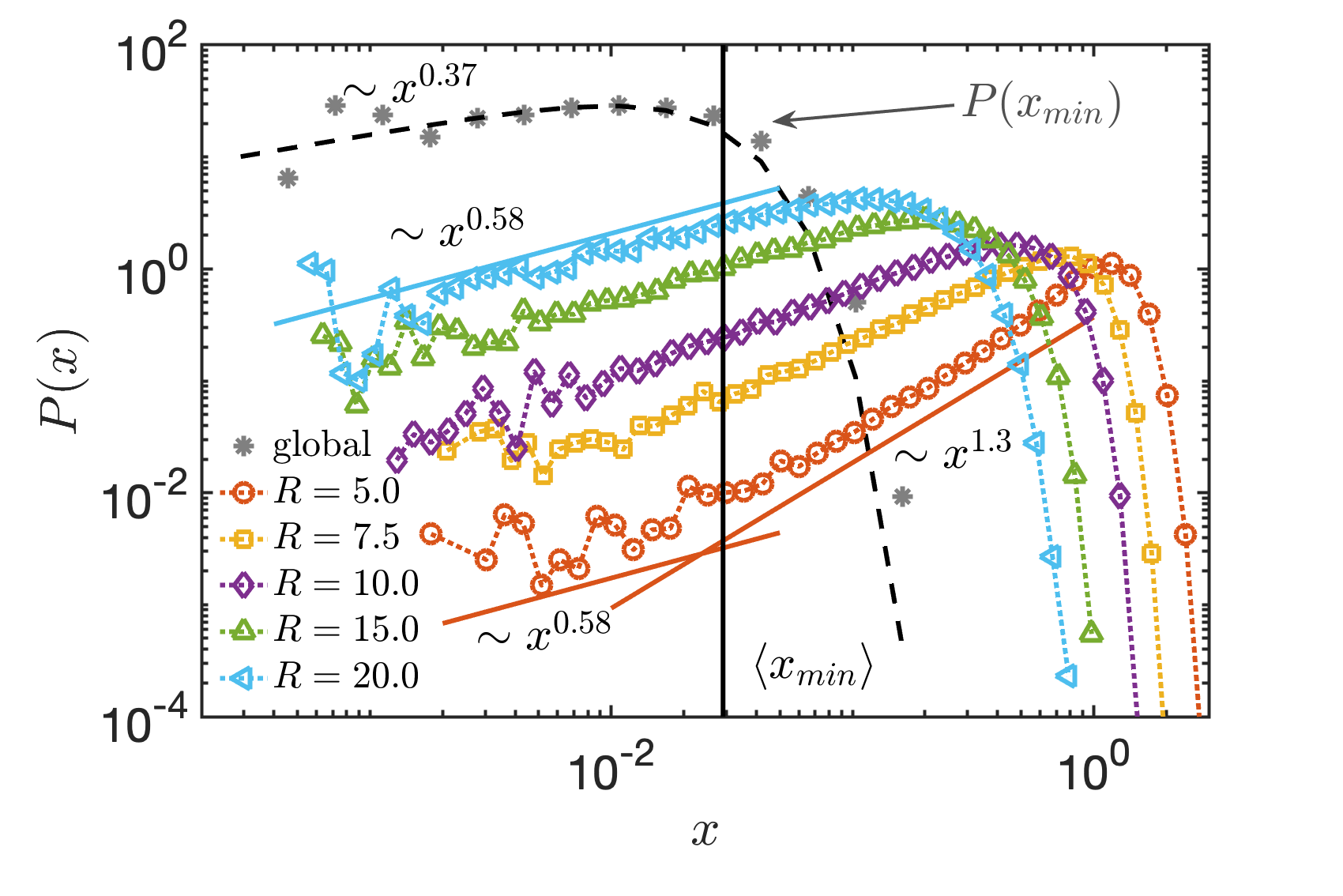}
\includegraphics[scale=0.12]{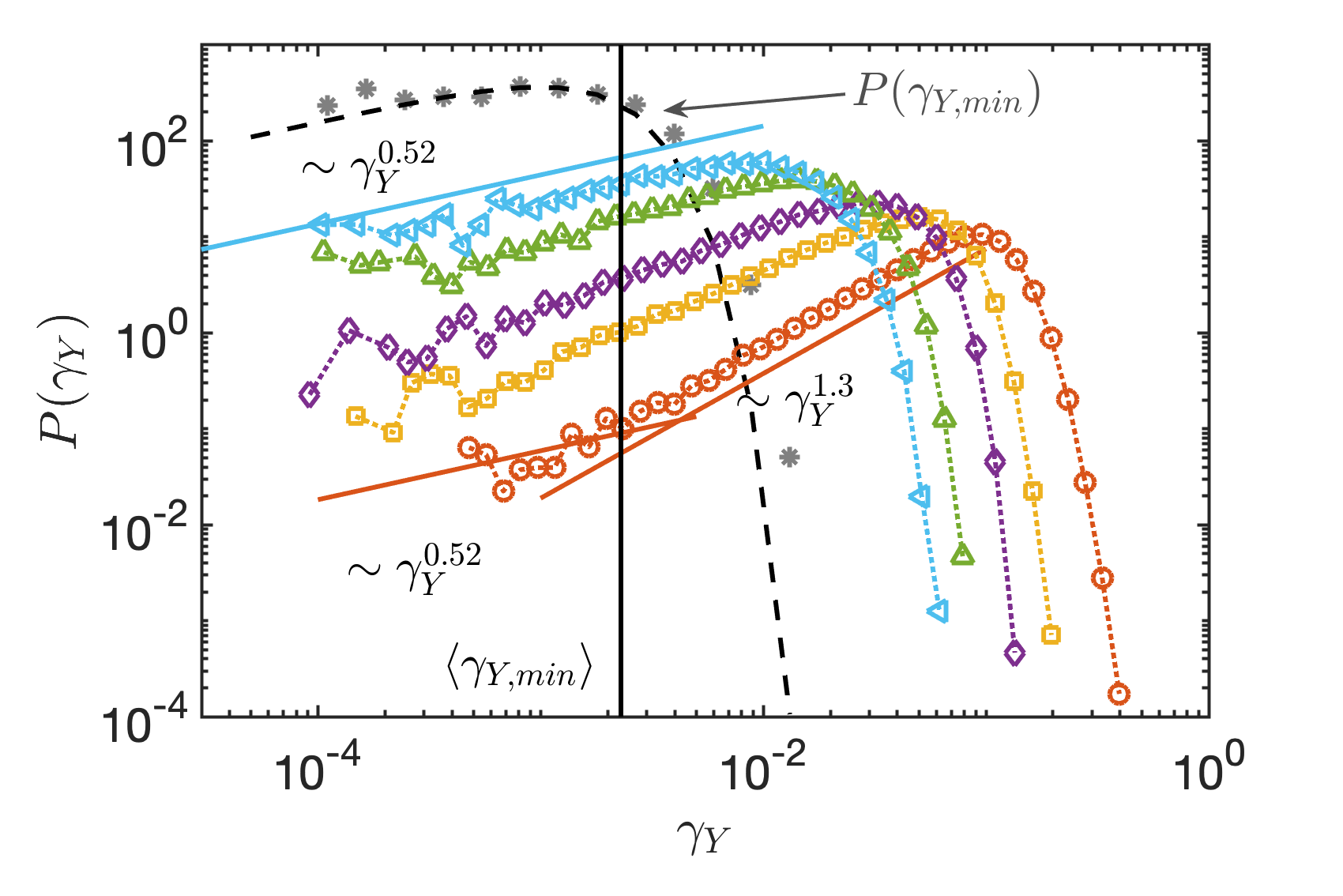}
\includegraphics[scale=0.22]{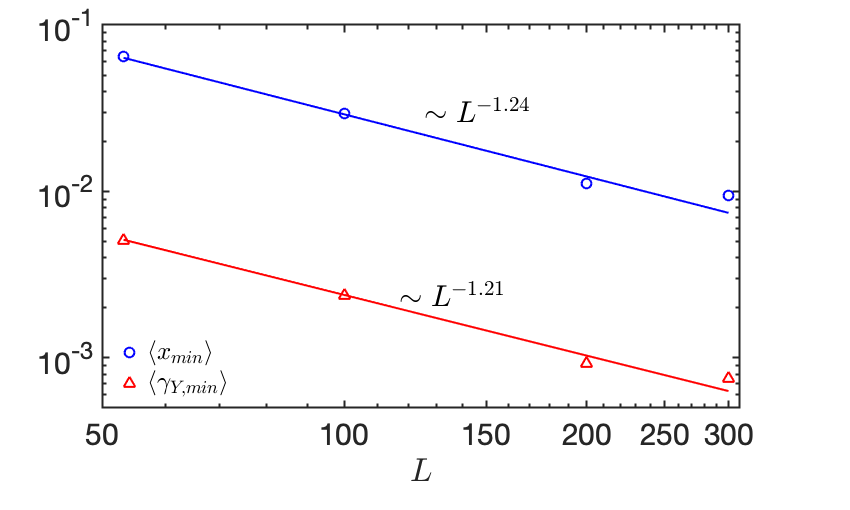}
\caption{Comparison between the distribution of weakest sites $P(y_{min})$ $(\star)$ computed from the global configurations and the distribution of $P(y)$ (open symbols) obtained by the local FM method for circular regions of different sizes $R$. The top panel shows results for the residual stresses $y=x$ while the middle panel shows the yield strains $y=\gamma_Y$. In both panels, the dashed lines correspond to Weibull fits to the global data. The bottom panel shows the scaling of $\langle x_{min} \rangle$ and $\langle \gamma_{Y,min} \rangle$ with system size $L$. }
\label{fig:pdf_quench}
\end{figure}

\begin{figure*}[t]
\centering
\includegraphics[scale=0.32]{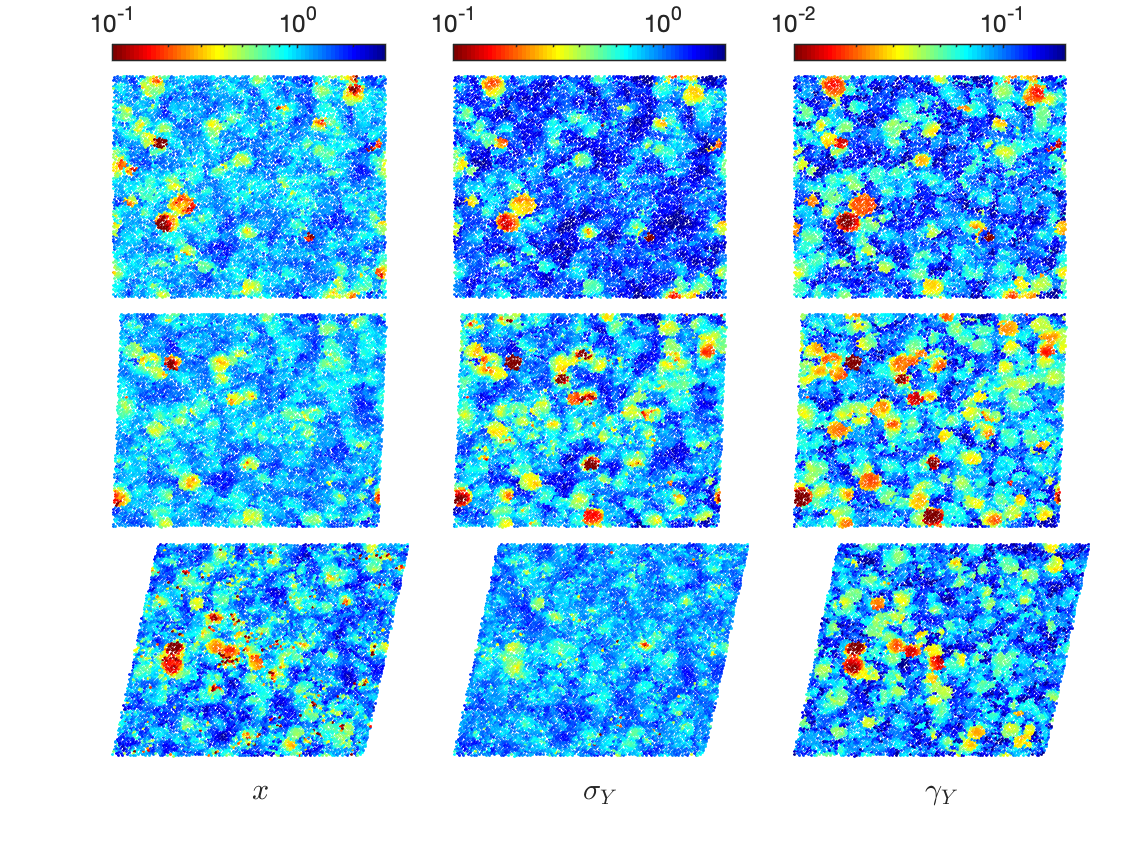}
\caption{Maps of residual stress $x$, local yield stress $\sigma_Y$,    and yield strain $\gamma_Y$ upon deformation for an applied strain $\gamma=0$ (Top), $\gamma=0.03$ (Middle) and $\gamma=0.18$ (Bottom). }
\label{fig:map}
\end{figure*}

In Figure \ref{fig:pdf_quench} we show $P(x)$ and $P(x_{min})$ (Top) and $P(\gamma_Y)$ and $P(\gamma_{Y, min})$ (Middle). We first observe that the Weibull distribution, when fitted to our data, describes well the distributions $P(x_{min})$ and $P(\gamma_{Y, min})$  coming from the global configurations.  However, we notice that the Weibull exponent $\theta=0.37$ is smaller in the case of weakest residual stresses than in the case of smallest yield strain, where we obtain $\theta=0.52$. We attribute the difference to the limited statistics obtained for both quantities in the very small $x_{min}$ and $\gamma_{Y,min}$ regions. 

Inspection of the distributions $P(y;R)$ reveals the effects of size $R$ and of the constraint imposed by the frozen shell. For both residual stresses and yield strains, increasing $R$ induces a shift of the distribution toward smaller values of $y$.
However, the distributions show a pseudogap form at small $x$ and rapidly decaying part at larger $x$. For $R < 10$,  we notice the presence of two power-law regimes. One regime at very small values exhibits an exponent $\theta_1=0.58$ for $P(x)$ which is larger than the exponent $\theta= 0.37$ extracted from $P(x_{min})$. For the yield strain, however, $\theta_1=0.52$ is in perfect agreement with $\theta$ extracted from the Weibull fit of $P(\gamma_{Y,min})$ and also in good agreement with $\theta_1$ from the residual stress. 

For larger values of $y$ and $R=5.00$, there appears to be a transition to a steeper power-law regime where $\theta_2=1.3$. This second regime is a consequence of the frozen boundary, which prevents nonaffine relaxations outside the circular region and therefore makes the rearrangements more difficult. This effect is particularly important for small local regions but, as $R$ increases, relaxation becomes easier and the two power law regimes merge into one unique regime when $R \ge 10$. 

For the weakest sites which are on the verge of yielding, the rearrangements do not require a substantial deformation and the constraints induced by FM do not affect their relaxation. Therefore, the distributions obtained in small $x$ or small $\gamma_{Y}$ regions show what we believe to be the right pseudogap form. To confirm our hypothesis, we look at the scaling of  $\langle x_{min} \rangle$ and $\langle \gamma_{Y,min} \rangle$ with the system size in the global unconstrained configurations. Results are shown in the bottom panel of Figure \ref{fig:pdf_quench}, where we see that $\langle y_{min} \rangle \sim L^{-\alpha}$ with $\alpha=1.24$ in the case of $\langle x_{min} \rangle$ and $\alpha=1.21$ in the case of $\langle \gamma_{Y,min} \rangle$. Using extreme value statistics, $\theta=2/\alpha -1$ (see also eq.~(\ref{eq_cdf})), we thus find $\theta=0.61$ (resp. $\theta=0.65$) for residual stresses (resp.~yield strains). These values are in good agreement with $\theta_1$ extracted from the distributions obtained with FM. The results obtained in the quenched state are in agreement with what has been reported by other atomistic simulations that found $\theta \approx 0.6$ for deformations on global systems \cite{Karmakar2010Rapid,Hentschel2015}. They are also coherent with results obtained from elastoplastic mesoscopic and mean-field models, for which $P(x)$ can be determined easily, which find $\theta \approx 0.6$ \cite{LinWyartEPL} and  $\theta \approx 0.5$ \cite{LinWyartPRX} respectively.

As the FM method is able to capture correctly the pseudogap exponent in the quenched state, we can use it to probe what occurs during deformation. As shown before, the shape of $P(x)$ is strongly influenced by the size of the circular region. One possibility is to consider local regions of $R \ge 10$ to avoid dealing with the two power law regimes, but then each local region would include more than 300 particles. Recent work from  Barbot {\it{et al.}} shows a good correlation between local yield stress and plastic activity for $R=5.0$ $(\sim 80$ particles) and thus argues for using a smaller size of system to correctly capture plastic activity. This value also coincides with the lower bound for which Hooke's law still applies \cite{Tsamados2009}. For these reasons, in what follows, we set the size of the frozen region to $R=5.0$.

\section{Evolution upon deformation}

\subsection{Local plastic observables}
In order to probe the effect of deformation on the local plastic observables $x$, $\sigma_Y$ and $\gamma_Y$,  we consider $N_g$ different global configurations of size $L$ that we deform up to $20 \%$. Snapshots are saved every $1\%$ of deformation in the transient regime and every $2\%$ of deformation in the stationary regime. As for the quenched state, we consider $N_{\ell}$ independent local regions of size $R=5.0$ selected from global configurations saved at given applied strain $\gamma$. 

Examples of maps of these quantities for several values of applied strain can be seen in Figure \ref{fig:map}.  As also found by Patinet {\it{et al.}}, \cite{Patinet2016,BarbotPatinet2018}, heterogeneities are well-visible in the quenched state for residual stress, yield stress and yield strain. We see the presence of weak spots embedded into an harder medium. The maps of the residual stress and yield strain are very similar due to the nature of the loading protocol. Indeed less deformation is required to make weak sites fails, so sites with small $x$ are also sites with small $\gamma_Y$.

\begin{figure}[t]
\centering
\includegraphics[scale=0.12]{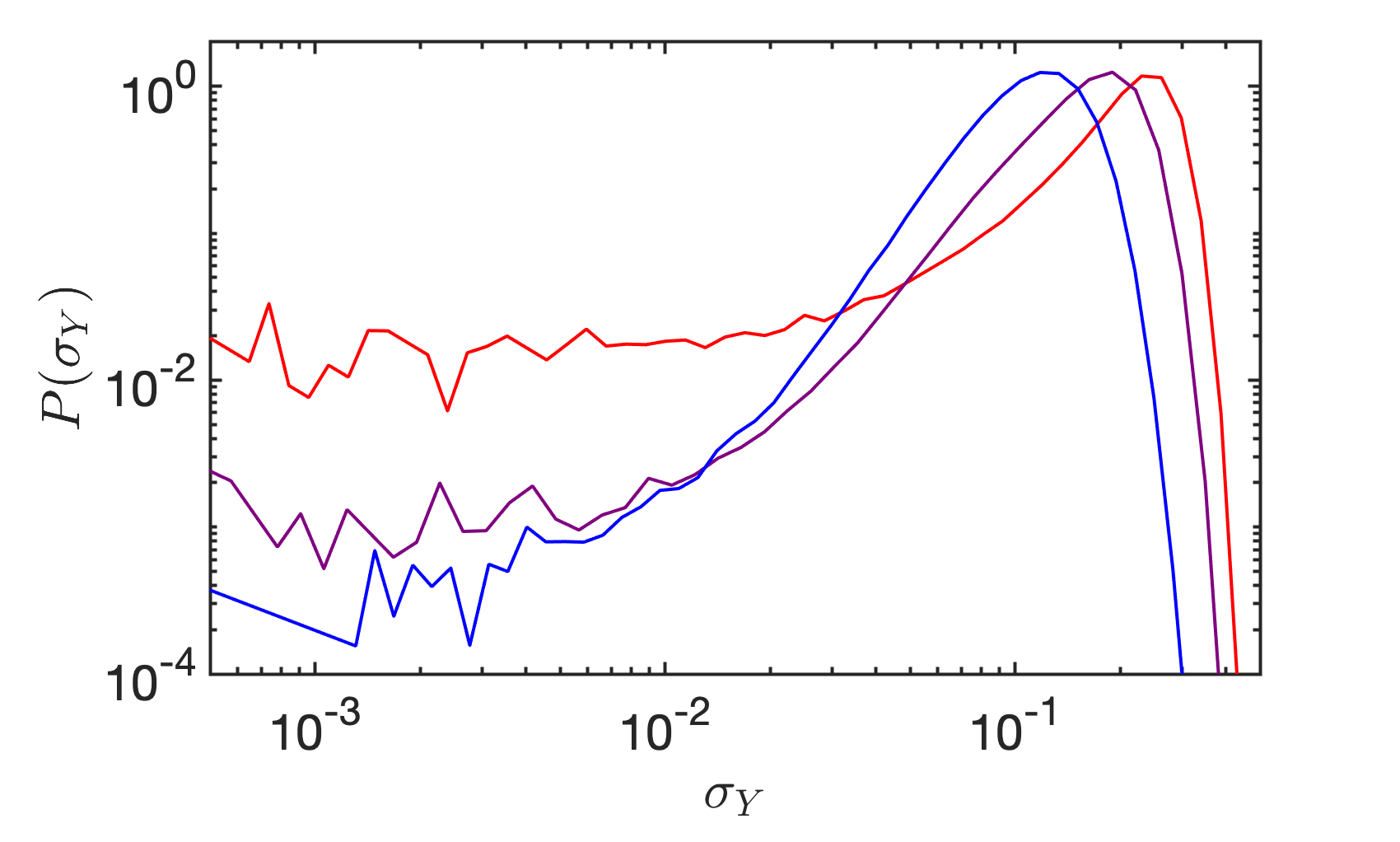}
\includegraphics[scale=0.12]{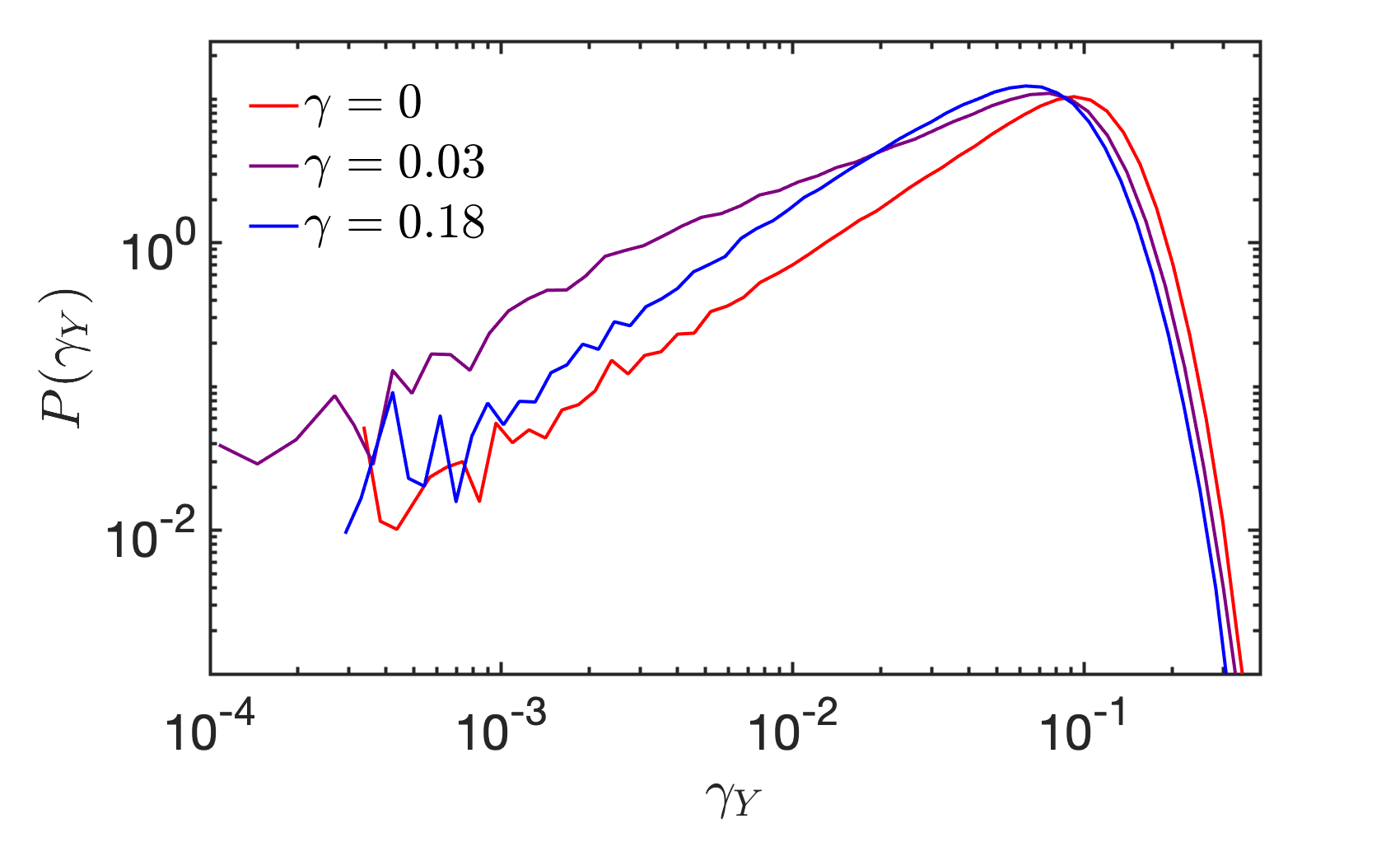}
\caption{Probability distributions of the yield stress, $\sigma_Y$ (Top) and of the yield strain, $\gamma_{Y}$ (Bottom) for three applied strains in the quench, transient and stationary regimes.}
\label{fig:evolution_distri}
\end{figure}

\begin{figure}[t]
\centering
\includegraphics[scale=0.28]{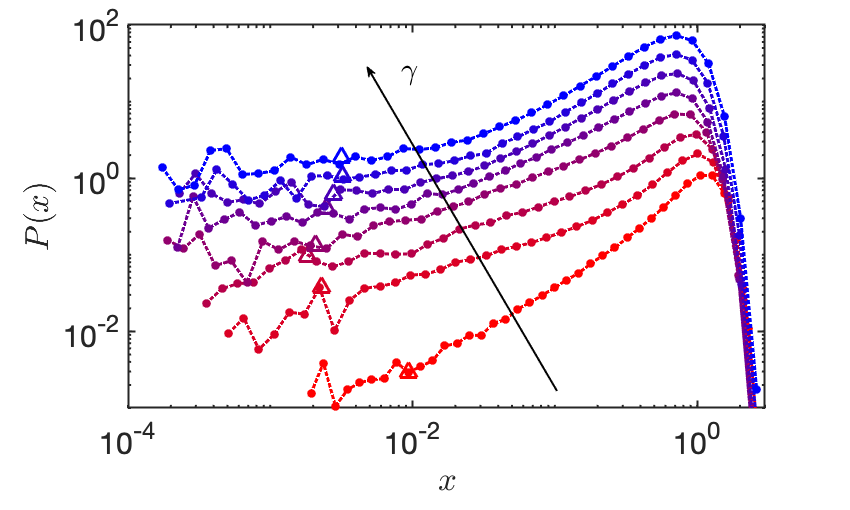}
\caption{Evolution of the distribution of residual stresses $P(x)$ with increasing applied strain $\gamma=0,0,01,0.02,0.04,0.06,0.08,0.12,0.18$ (from top to bottom) in systems of size $L=300$. The triangles represent the location of $\langle x_{min} \rangle$ obtained from global deformation of the configurations. The different curves have been shifted in the upper direction for sake of clarity.}
\label{fig:evolution_pdf}
\end{figure}
 
Upon deformation, we notice that the number of sites with very weak values of $\sigma_Y$ decreases significantly pointing toward a homogenization of the system in the stationary regime. This observation is also visible in the top panel of Figure \ref{fig:evolution_distri} where the probability of the local yield stress, $P(\sigma_Y)$, is smaller at small $\sigma_Y$ in the transient and stationary regimes than in the quenched state. Moreover, there appears to be a tendency of hard spots to become softer in the transient regime. Maps in the transient regime show the larger number of weak spots as we also observe in the probability distribution function (pdf) of the yield strain $P(\gamma_Y)$ displayed in the bottom panel of Figure \ref{fig:evolution_distri}. Indeed the fraction of small $\gamma_Y$ is larger in the transient regime than in the quenched and stationary regime which have almost the same fraction of small $\gamma_Y$.

\subsection{Plateaus and mechanical noise}

\begin{figure}[t]
\centering
\includegraphics[scale=0.26]{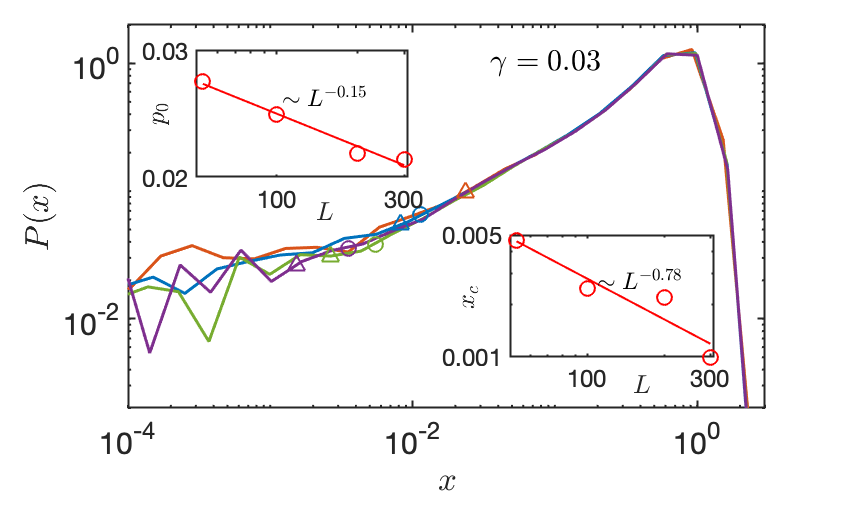}
\includegraphics[scale=0.26]{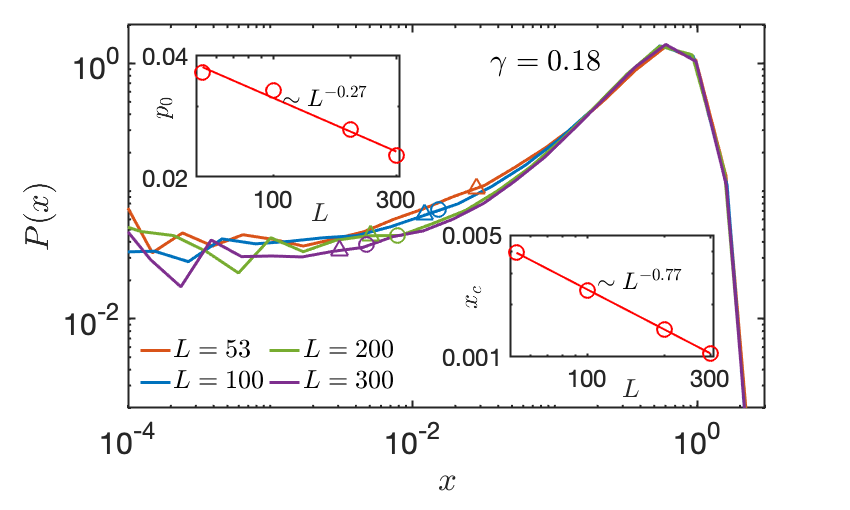}
\caption{Effect of the system size $L$ on $P(x)$ in the transient (Top) and stationary regime (Bottom). The triangles indicate the location of $\langle x_{min} \rangle$  while the circles stand for the location of the lower cutoff of the mechanical noise $\Delta x$, see Fig.~\ref{fig:noise}. The insets shows the scaling with $L$ of the plateau $p_0$ and the crossover $x_c$ from the power law regime to the plateau. }
\label{fig:pdf_size_scaling}
\end{figure}

\begin{figure}[t]
\centering
\includegraphics[scale=0.12]{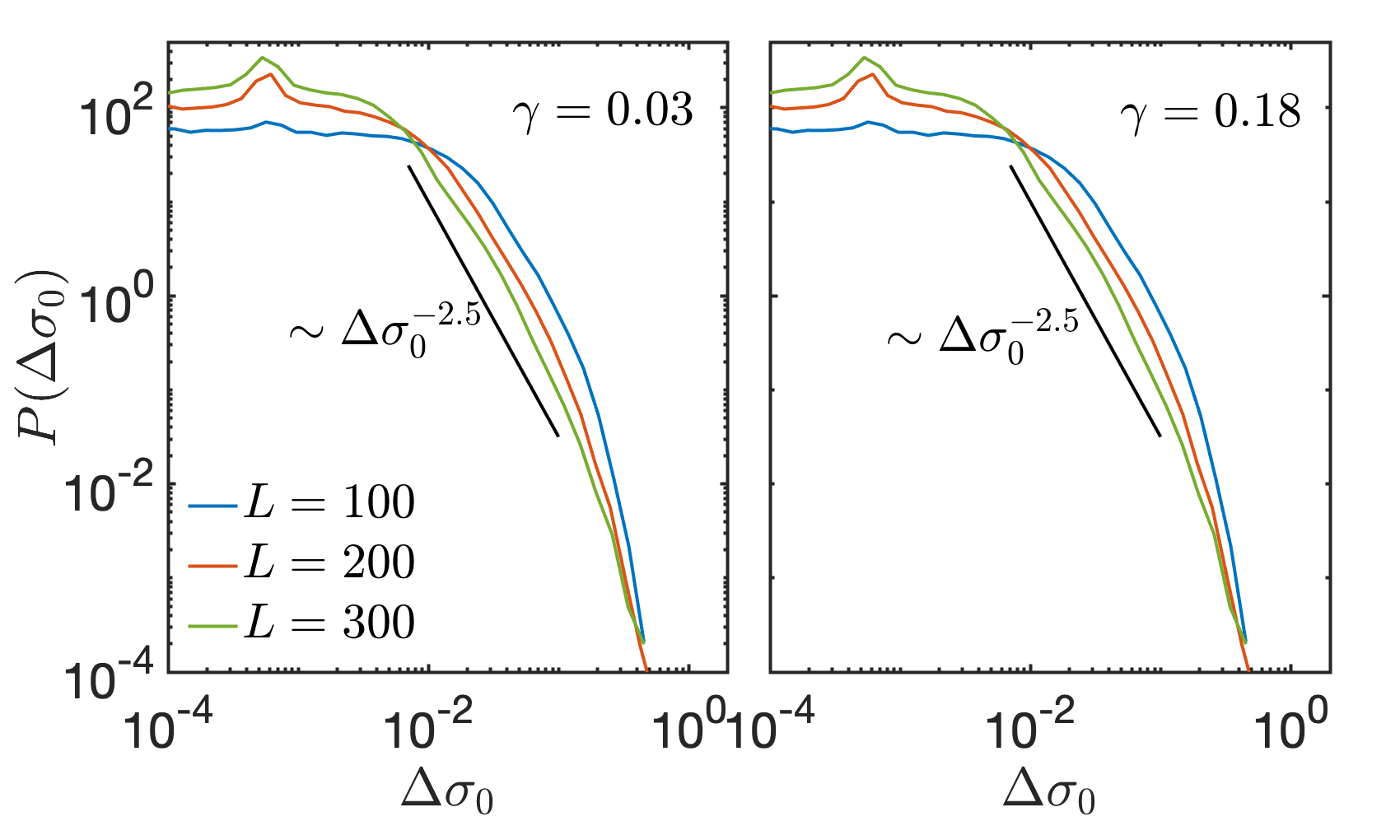}
\includegraphics[scale=0.12]{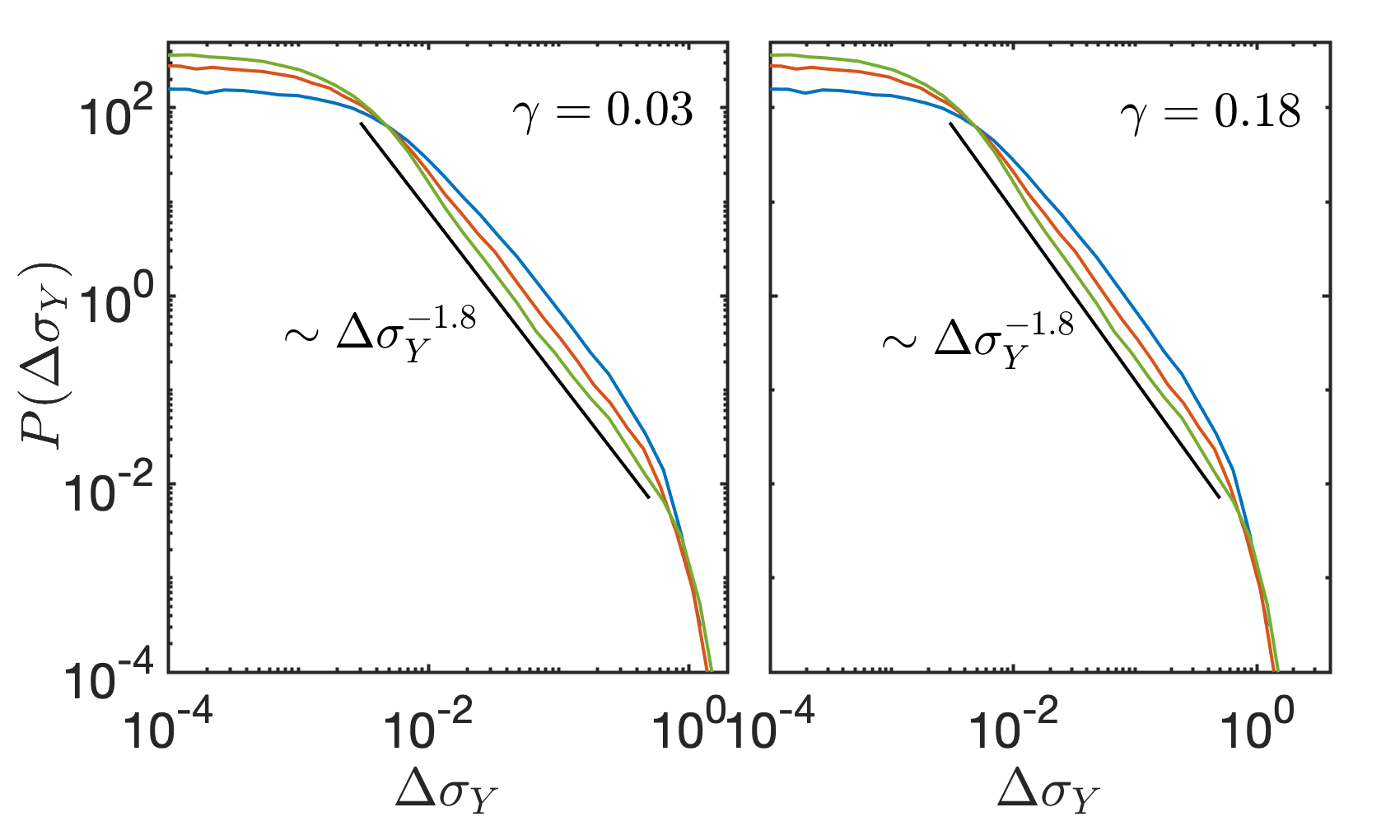}
\includegraphics[scale=0.12]{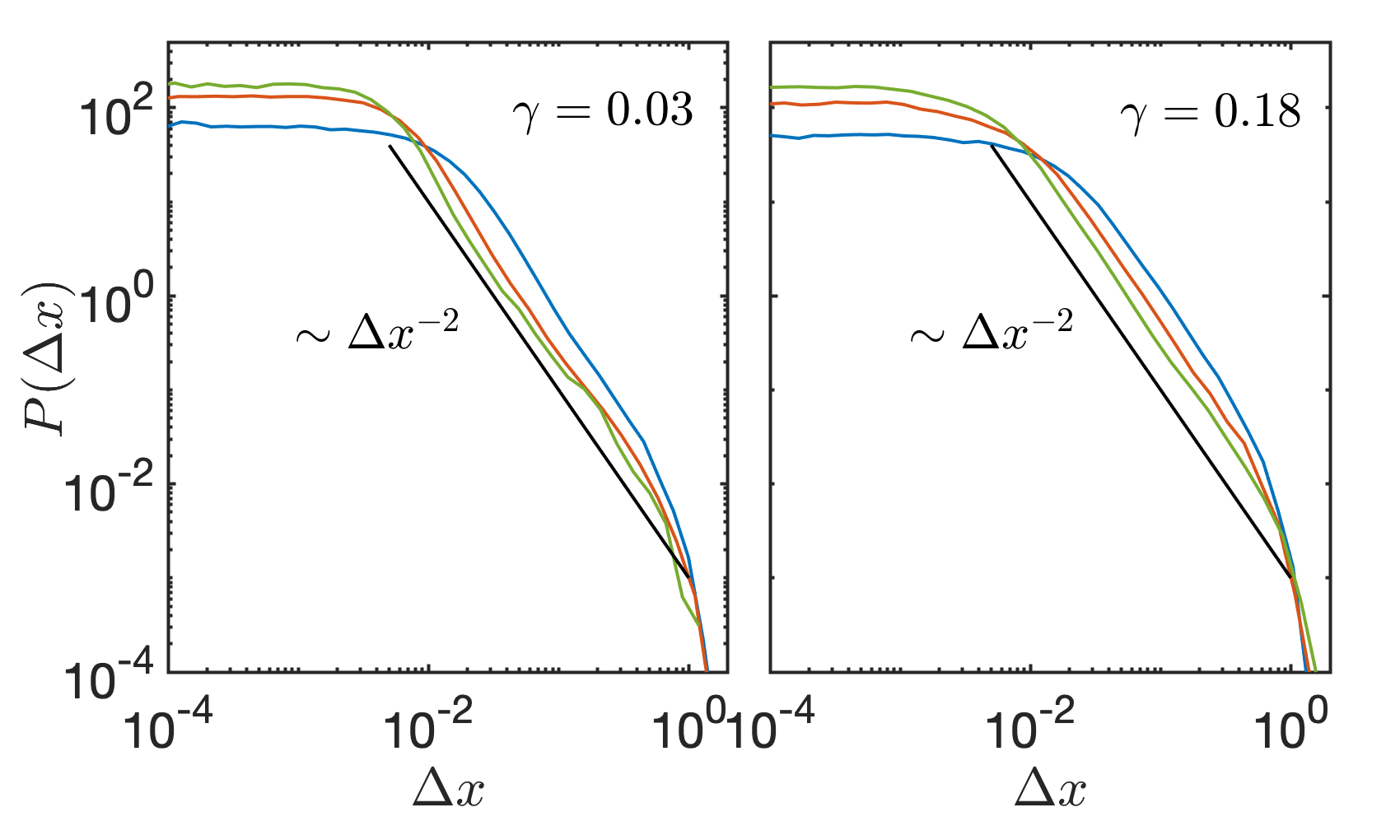}
\includegraphics[scale=0.24]{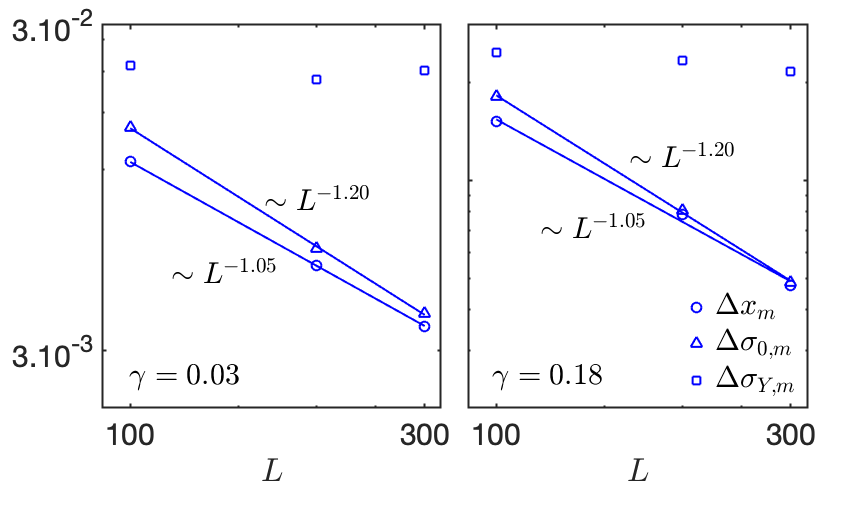}
\caption{Distribution of the changes in local stress $\Delta \sigma_0$, yield stress $\Delta \sigma_Y$ and residual stress $\Delta x$ between consecutive avalanches in the transient (left) and stationary (right) regimes. The bottom panel shows the system size scaling of the lower cutoffs of these distributions. }
\label{fig:noise}
\end{figure}

The evolution of residual stresses upon deformation is shown in Figure \ref{fig:evolution_pdf}. At the earliest stages of deformation, we find $P(x) \sim x^{\theta}$ in the small $x$ region and the pseudogap exponent $\theta$ is decreasing.  After only $3\%$ strain, we notice the appearance of a plateau which remains visible for larger applied strains. Such deviations from the pure pseudogap form of $P(x)$ in the steady state have recently been reported in two different works that employed coarse-grained idealized elastoplastic models (EPM) \cite{Tyukodi2019,Ferrero2019}. These observations could be described by rewriting the pdf of residual stresses as $P(x) \sim p_0 + x^{\theta}$, where $p_0 \sim L^{-p}$. A system size dependence of the plateau value of $P(x)$ also appears in our atomistic simulations as reported in Figure \ref{fig:pdf_size_scaling}, where $P(x)$ are shown for different sizes $L$ and for two given applied strains, $\gamma=0.03$ (transient) and $\gamma=0.18$ (stationary).
In order to investigate the size scaling of the plateau value $p_0$ (see inset),  we average the value of $P(x)$ for $x \le 10^{-3}$ and find the plateau $p_0 \sim L^{-0.15}$ in the transient regime and $p_0 \sim L^{-0.27}$ in the steady-state regime. The values of these exponents are significantly lower than those reported in the EPM studies \cite{Tyukodi2019,Ferrero2019}.  In qualitative agreement with these studies, however, we observe that when the system size $L$ increases, the crossover from the power law region to the plateau occurs at smaller values of $x$. We estimate the value $x_c$ at which the crossover takes place by looking at the intersection between the plateau and the power law region. As shown in the inset of Figure \ref{fig:pdf_size_scaling}, in the stationary regime we find that $x_c \sim L^{-0.78}$. This value is in agreement with EPM results for which the exponent has been estimated between 0.73-0.95 \cite{Tyukodi2019, Ferrero2019}. 

\begin{figure}[t]
\centering
\includegraphics[scale=0.28]{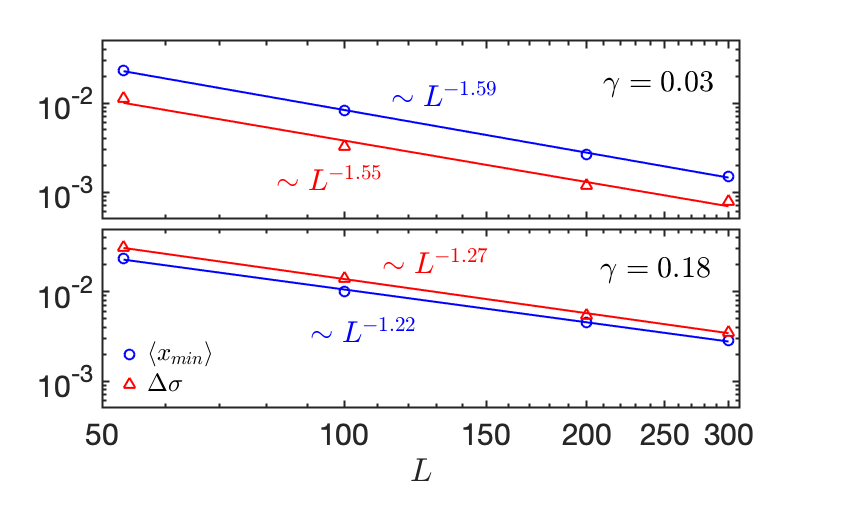}
\includegraphics[scale=0.25]{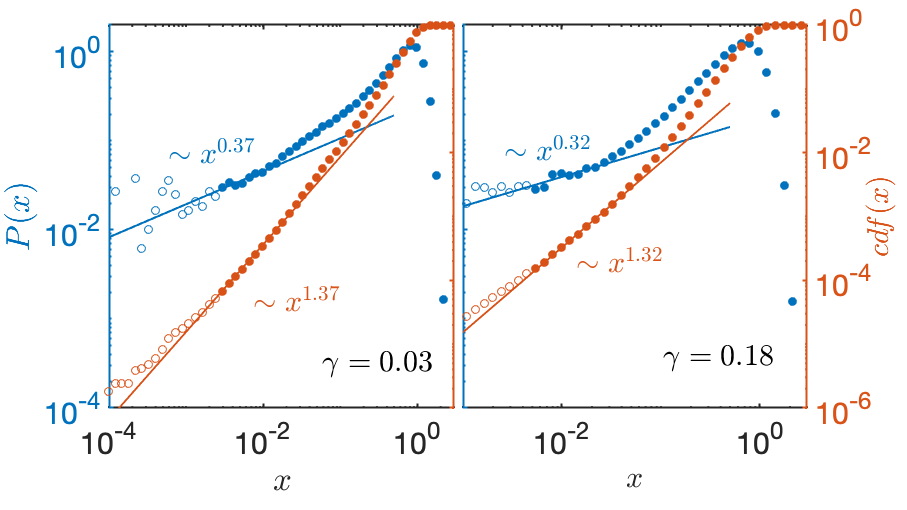}
\caption{Upper panel: System size scaling of $\langle x_{min} \rangle$ and $\Delta \sigma$ in the transient (Top) and stationary (Bottom) regimes. Lower panel: Examples of fits performed on $P(x)$ and $cdf(x)$ in the transient and stationary state for $L=300$. The filled symbols show the range of $x$ values for which $x > \Delta x_c$. }
\label{fig:alpha_computation}
\end{figure}

The origin of the plateau can be attributed to the discreteness of the underlying stochastic process as suggested by Zoia et al.~\cite{Zoia} in the context of fractional Brownian motion. The dynamics of the variable $x$ can be viewed as a random walk in the presence of an absorbing boundary at $x=0$. One expects deviations from scale-free behavior of the steady state distribution $P(x)$ near that boundary when $x$ becomes of the order of the typical increment of the mechanical noise coming from the stochastic redistribution of the stress during plastic rearrangements. The plateau is therefore a finite size effect that vanishes in the thermodynamic limit. The distribution of stress kicks $\Delta \sigma_0$, which monitors the variation of stress between two consecutive plastic rearrangements, is known to be broadly distributed and can be described by a truncated Pareto distribution \cite{LinWyartPRX}, 
\begin{align}
\rho(\Delta \sigma_0) = \frac{\mu \Delta \sigma_{0,c}^{\mu}}{|\Delta \sigma_0|^{1+\mu}}
\label{eq:noise}
\end{align}
where $\Delta \sigma_{0,c} \sim L^{-2/\mu}$ is the lower cutoff of the distribution which depends on the system size and sets a characteristic scale. The distribution also has an upper (system size independent) cutoff which comes from the variation of stress felt by adjacent sites. The exponent $\mu$ is related to the elastic interaction kernel $\mathcal{G}(r) \sim r^{-d/\mu}$. For $\mu=1$, we recover the Eshelby kernel \cite{LinWyartPRX} that can be well fit to elementary shear transformations observed in atomistic glass models \cite{albaret2016}. 

The AQS scheme allows to resolve individual avalanches but cannot discriminate between single and multiple plastic events inside a given avalanche. Therefore, noise kicks are defined between the end points of consecutive avalanches. In the transient and the stationary regimes, we record all stress drops in an applied strain interval of $2 \%$. From our data we can determine not only the distribution of stress kicks $P(\Delta \sigma_0)$ between two consecutive avalanches, but also the distribution of the variation of yield stress $P(\Delta  \sigma_Y)$ and the distribution of the variation of the residual stress $P(\Delta x)$ for the sites that did not fail during theses avalanches. The results are shown in Figure \ref{fig:noise}, where we observe that the stress kicks obey $P(|\Delta  \sigma_0|) \sim |\Delta  \sigma_0|^{-2.5}$ with a system size dependent lower cutoff (defined by the intersection of the flat region of the distributions with the power law region) that varies as $L^{-1.2}$ (Bottom panel). The scaling of the lower cutoff implies a slightly overestimated value of $\mu=2/1.2\approx 1.67$, that could be related to the fact that we estimate the scaling on small interval \cite{info_noise}. 
The small peaks in the flat region come from the fixed strain increment as a typical step generates a stress $\sim G \delta \gamma \approx 1.2 \cdot 10^{-3}$. The data is thus consistent with eq.~(\ref{eq:noise}) with $\mu=1.5$, which is a larger exponent than what is expected for an Eshelby kernel. Within the AQS protocol, we can only probe stress changes between entire avalanches and not single plastic rearrangements. Therefore, the stress kicks that we measure are not coming from spatially localized events but rather from extended ones, for which $\mu>1$ is expected \cite{Jagla2018MC}. 

Interestingly, the middle panel of Fig.~\ref{fig:noise} shows that the local yield stress also changes even if the region did not undergo a shear rearrangement. The atomistic simulations are thus at variance with EPMs, which usually assume $\Delta \sigma_Y=0$, and the distribution of kicks $P(\Delta  \sigma_0)$ is equivalent to the distribution $P(\Delta  x)$. In our system, the yield strain on stable sites varies between two consecutive rearrangements as $P(|\Delta  \sigma_Y|) \sim |\Delta  \sigma_Y|^{-1.8}$. Even if the local environment of particles  in terms of nearest neighbors does not change, long ranged nonaffine displacements that occur (even during the elastic branches of loading) perturb the local environment. Already small nonaffine changes in the atomic positions are likely to be sufficient to slightly modify the local yield stress.

Coming now to the distribution of the variation of the residual stress itself, the bottom panel of Fig.~\ref{fig:noise} shows that $P(|\Delta x|) \sim |\Delta x|^{-2}$, implying $\mu=1$ and therefore a compatibility with the Eshelby kernel. However, we believe this value to be coincidental. Moreover, the system size scaling of the lower cutoff $\Delta x_c \sim L^{-1.05}$ is not consistent with eq.~(\ref{eq:noise}). We attribute these differences to the contributions $\Delta \sigma_Y$ from the changing yield stress to $\Delta x$. We also note that the scaling of $\Delta x_c$ is reasoably close to the scaling that we found for $x_c$.

The location of the lower cutoff  $\Delta x_c$ on the distribution $P(x)$ is indicated with open circles in Figure \ref{fig:pdf_size_scaling}. We observe that for the largest system sizes $L=200$ and $L=300$, $\Delta x_c$ marks the entrance of the plateau region, while for $L=100$, it appears to be still in the power law region. Given the limitations of accessible system sizes with the atomistic simulations, these observations are consistent with the notion expressed above that $\Delta x_c$ is the relevant discretization scale below which deviations from the  pseudogap form of $P(x)$ become visible.

\subsection{Scaling relations}

\begin{figure}[t]
\centering
\includegraphics[scale=0.28]{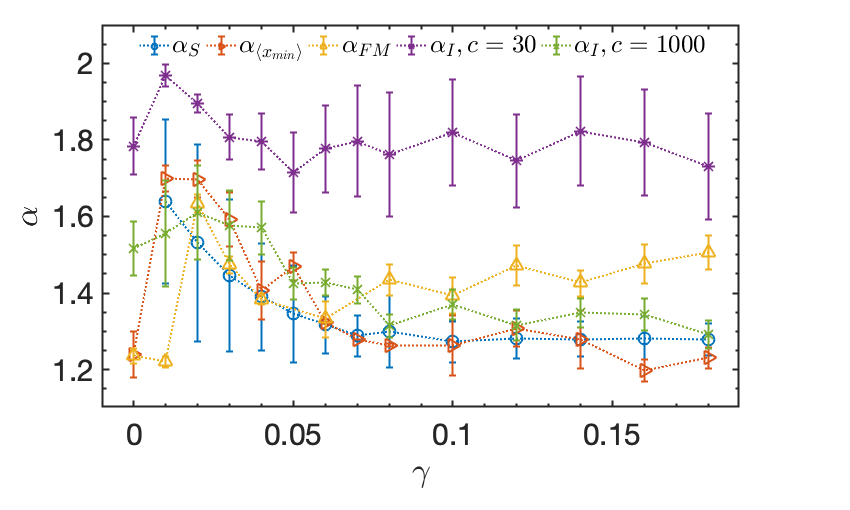}
\caption{Evolution of the exponent $\alpha$ with applied strain $\gamma$ computed from 4 different methods: scaling of the stress drops, $\alpha_S$, scaling of the weakest sites, $\alpha_{\langle x_{min} \rangle}$, from the pseudogap exponent $\theta$ extracted from $P(x)$ computed from the FM, $\alpha_{FM}$, and from the scaling of $\langle x_{min} \rangle$ obtained from the inversion of the cdf, $\alpha_I$. }
\label{fig:alpha}
\end{figure}

The weakest residual stresses are controlling of the flow of amorphous solids, which in the athermal quasistatic limit consists of periods of elastic loading punctuated by sudden stress drops. The magnitude of these stress drops decreases with increasing system size as $\langle \Delta \sigma \rangle \sim L^{-\alpha_S}$. In the upper panel of Fig.~\ref{fig:alpha_computation} we compute the average size of the stress drops induced by global deformation vs.~$L$ and find a steady state value $\alpha_S=1.27$ in very good agreement with previous atomistic simulations \cite{SalernoRobbins,LiuBarrat2016}. In the transient regime, a larger value $\alpha_S=1.55$ is found. Also shown is the behavior of $\langle x_{min} \rangle$ in these simulations, which also follows a scaling form $\langle x_{min} \rangle \sim L^{-\alpha_{\langle x_{min} \rangle}}$. If one assumes that on average loads and releases compensate, one expects 
$ \langle \Delta \sigma \rangle \sim \langle x_{min} \rangle$. Figure \ref{fig:alpha} reports that these two exponents agree well with each other for the entire range of deformation strain. As soon as deformation sets in, the exponents jump from a (preparation dependent) initial value near 1.2 to a larger value of 1.7 and then decrease during the transient loading phase to a steady state value 1.27. Similar trends have been observed in prior atomistic simulation studies \cite{Hentschel2015,OzawaPNAS}.

The mean value of the weakest site $\langle x_{min} \rangle$ can be computed from the underlying distribution of residual stresses via a well-known expression from extreme value statistics \cite{Karmakar2010Rapid}, 
\begin{align}
\label{eq_cdf} 
\int_0^{\langle x_{min} \rangle} P(x) dx = \frac{c}{L^d}\,\,  \Rightarrow \langle x_{min} \rangle \sim L^{-\alpha_{\langle x_{min} \rangle}}
\end{align}
From eq.~(\ref{eq_cdf}) we see that depending on the form of $P(x)$, $\alpha_{\langle x_{min} \rangle}$ can have different expressions. For instance, if we assume that $x  \le  \langle x_{min} \rangle$ is inside the plateau region then $P(x) \sim L^{-p}$ and $\langle x_{min} \rangle \sim L^{-(d-p)}$. However, if we assume that the pseudogap description $P(x) \sim x^{\theta}$ is valid then $\langle x_{min} \rangle \sim L^{-d/(1+\theta)}$. 

In what follows, we seek to ascertain if the distribution $P(x)$ that we obtained with the FM method is compatible with the exponents $\alpha_S$ and $\alpha_{\langle x_{min} \rangle}$ that were obtained from unconstrained global AQS deformations. In order to distinguish from these exponents,  we denote by $\alpha_{FM}$ the value of the exponents obtained from FM. From the observations made in Figure \ref{fig:pdf_size_scaling}, we see that $\langle x_{min} \rangle$ is located in the plateau region, which suggests $\alpha_{FM}=d-p$. In the transient regime $\alpha_{FM}=1.85$ while in the stationary regime $\alpha_{FM}=1.73$. These two values are obviously too large to be in agreement with what has been measured from global configurations. 

Alternatively, we might assume that the global scaling on $\langle x_{min} \rangle$ reflects the thermodynamic limit, where the plateau is irrelevant.  A careful inspection of $P(x)$ reveals that the crossover between the two power law regimes observed in Figure \ref{fig:pdf_quench} remains upon deformation. We can therefore locate the transition from a region where the power law is dominated by the FM artefacts coming from the truncation of nonaffine displacements to a region where we assume the pseudogap exponent to be valid. As $\langle x_{min} \rangle < \Delta x_c$ and is located either in the plateau or near the crossover region, we focus on the power law region for $x \ge \Delta x_c$.   Moreover, as noted before, the crossover toward the plateau decreases with increasing $L$. Therefore, we compute the cumulative distribution function (cdf) of $P(x)$ for $L=300$ and  consider only the range of $x$ values between the crossover to the steeper power law and the entrance to the plateau  ($x \ge  \Delta x_c$) as shown in the bottom panel of Figure \ref{fig:alpha_computation}. Once the pseudogap exponent $\theta_{FM}$ is found, we determine the scaling exponent as $\alpha_{FM}=d/(1+\theta_{FM})$. In Figure \ref{fig:alpha} we see that the agreement between $\alpha_{FM}$, $\alpha_S$ and $\alpha_{\langle x_{min} \rangle}$ is good in the transient regime but $\alpha_{FM}$ is slightly overestimated in the stationary regime. As mentioned above, the discreteness of the mechanical noise and the constraints imposed by the FM reduce significantly the range where $P(x) \sim x^{\theta_{FM}}$ can be observed.

Given the difficulties to extract a reliable value of $\theta_{FM}$ directly from fits to $P(x)$, we explore another approach, which considers eq.~(\ref{eq_cdf}) and extracts $\langle x_{min} \rangle$ from $P(x)$ by inverting its cdf, i.~e.~$\langle x_{min} \rangle = cdf^{-1}(cL^d)$. The resulting exponent is denoted $\alpha_{I}$.  The only free parameter in this procedure is the proportionality constant $c$. The choice of $c$ allows to explore regions of different values of $x$. We selected two values: $c=30$, which explores regions where $x \lesssim 0.01$ and  $c=1000$, for which $x \lesssim 0.1$ for the smallest system size $L=53$. The former range covers the values of $x$ in the plateau regions. In Figure \ref{fig:alpha}, we show the evolution of $\alpha_{I}$ upon deformation. For $c=30$, $\alpha_I$ is much larger than $\alpha_S$, $\alpha_{\langle x_{min} \rangle}$ and even $\alpha_{FM}$. For $\gamma \ge 0.03$, the scaling only reflects the presence of the plateau as $\alpha_I$ for $\gamma=0.03$ and $\gamma=0.18$ are in agreement with the values $\alpha_{FM}=d-p$ computed above. The values for $\gamma <0.03$ are also overestimated despite the absence of plateaus, possibly due to poorer sampling of small $x$. For $c=1000$, however, we observe a good agreement between $\alpha_I$ and  $\alpha_S$ and  $\alpha_{\langle x_{min} \rangle}$ for the whole range of deformation. This suggests that relevant information is contained in the initial power law region beyond the plateau. 

\section{Conclusions}
Using the FM method, we have examined the distribution $P(x)$ of residual stresses (or local thresholds to mechanical instability) in athermal 2D amorphous solids. In the quiescent (freshly quenched state), the FM method reveals a power law form with two different exponents. In the limit of small $x$, $P(x)\sim x^\theta_1$ where $\theta_1<1$ agrees well with an independent estimate of the pseudogap exponent $\theta$ from extreme value statitics. The FM method then shows a second power law regime with exponent $\theta_2>1$. In this regime we believe the local yield stress to be overestimated due to incomplete relaxation in the probed region, and this effect could be more pronounced for larger yield stresses.

As soon as deformation sets in, $P(x)$ becomes analytic and develops a plateau as $x \rightarrow 0$. The FM results thus show that similar observations made with mesoscale EPMs are generic and extend to a more detailed atomistic model. In order to elucidate the origin of this plateau, we have examined $P(\Delta x)$, the distribution of the residual stress differences between two consecutive avalanches. This distribution has a power-law form with a system size dependent lower cutoff $\Delta x_c(L)$ that endows the stochastic process with a characteristic scale. We believe that the plateau appears for $x<\Delta x_c(L)$, which is consistent with our data in the limited range of system sizes $L$ accessible to us in atomistic simulations.

The computation of the pseudogap exponent $\theta$ from the obtained distributions $P(x)$ presents challenges. The crossover into the plateau region severely limits the region in which $\theta$ can be determined by direct fits to $P(x)$. Only larger system sizes can help. Moreover, even though the values of $\langle x_{min} \rangle$ obtained from unconstrained global simulations fall into the plateau region for our largest system sizes, the observed system size scaling is not consistent with that predicted from the plateau itself. One possibility is that in the FM calculation, the contribution from the changing local yield stress $\Delta \sigma_Y$ to the total change $\Delta x =  \Delta \sigma_Y-\Delta \sigma_0$ is somehow overestimated. The plateau thus appears sooner than in the actually sampled residual stress distributions.

One of the most intriguing observations in the present study is a changing local yield stress in regions not experiencing a plastic event. This possibility is not conceptualized in current mesocale elastoplastic \cite{nicolas2018deformation} or mean-field \cite{LinWyartPRX} models of amorphous plasticity. Further improvements to the FM method that reduce the boundary artefacts and allow a better relaxation are needed to further explore this additional physics at the atomistic level.

A lingering question remains: what is the significance of the existence of this plateau in $P(x)$ for the statistical properties of the yielding transition? We suggest that the answer depends on the precise form of the system size scaling of the characteristic scale of the mechanical noise increments $\Delta x_c(L)$. For ideal "Eshelby sources" and no changes of the local yield stress except upon yielding (as assumed in EPM and mean-field treatments), we have  $\Delta x_c(L)\sim L^{-2}$ from eq.~(\ref{eq:noise}). In this case, the plateau crossover decreases faster with $L$ than $\langle x_{min} \rangle\sim L^{-\alpha_{\langle x_{min} \rangle}}$ $(\alpha_{\langle x_{min} \rangle}<2)$ and $\langle x_{min} \rangle$ must remain in the pseudogap region. The resultant scaling relations that connect the exponent $\theta$ with the exponents $\tau$ and $d_f$ that describe the statistics of macroscopic avalanches remain unaltered \cite{LinWyart2014}. The present results, however, suggest at least the possibility that $\Delta x_c(L)$ vanishes slower than $\langle x_{min} \rangle$ and thus the scaling of $\langle x_{min} \rangle$ becomes dominated by
the plateau regime for large enough system sizes.  In this case, the scaling relationship that links the pseudogap exponent $\theta$ to the avalanche exponents $\tau$ and $d_f$ will be altered. Our present computational capabilities are insufficient to settle this question definitively.

\section{Acknowledgements}
We thank Peter Sollich for discussion and useful comments on our manuscript. 
J.R. thanks the Alexander von Humboldt Foundation for financial support.  High performance computing resources were provided by ComputeCanada and the Quantum Matter Institute at the University of British Columbia. This research was undertaken thanks, in part, to funding from the Canada First Research Excellence Fund, Quantum Materials and Future Technologies Program. C.R. is part of the ANR LatexDry project, grant ANR-18-CE06-0001 of the French Agence Nationale de la Recherche.

\section*{Appendix: Determination of plastic events}
To determine plastic events, we revisit a criterion proposed by Lerner and Procaccia \cite{Lerner2009}, which relies on the difference in potential energy between the affine deformation $U_{\text{aff}}$ and the underlying inherent structure $U_0$ after deformation:
\begin{align}
	\kappa =\frac{U_{\text{aff}}-U_0}{N\delta \gamma^2}
\end{align}
where $\delta \gamma$ is the strain increment. While in ref.~\cite{Lerner2009} a reasonable but arbitrary value of $\kappa$ was selected, we demonstrate here that there exists a range of $\kappa $ associated with purely elastic deformation and therefore a lower bound above which $\kappa$ captures plastic events only. 

The demonstration relies on the hypothesis that we work in the elastic regime. We look at the energy variation after an AQS step. The two stages in the AQS protocol lead to variation in the strain energy density that can be described as follow: 
\begin{enumerate}
\item During the affine deformation the strain energy density evolves with respect to the starting configuration as: 	
 \begin{align}
\rho U_{\text{aff}}	= \rho U_{IS} + \sigma_0 \delta \gamma + \frac{G_{B}}{2} (\delta \gamma)^2
\end{align}
where $\sigma_0$ and $U_{IS}$ are respectively the residual stress and the inherent structure energy associated to the configuration before a strain increment is applied. $G_{B}$ is the affine shear modulus. 
 \item After the relaxation the strain energy density with respect to starting configuration is given by:
 \begin{align}
\rho U_0	= \rho U_{IS} + \sigma_0 \delta \gamma + \frac{G}{2} (\delta \gamma)^2
\end{align}	
where $G$ is the shear modulus. 
\end{enumerate}

Consequently the difference in strain energy density gives:
\begin{align}
	\rho (U_{\text{aff}} - U_0) = \frac{G_{B} - G}{2}(\delta \gamma)^2
\end{align}

Finally, we can estimate an upper bound for $\kappa$ in the elastic regime:
\begin{align}
\kappa=  \frac{G_{B}-G}{2 \rho}  \le \frac{G_B}{2\rho}
\end{align}

\begin{figure}[t]
\begin{center}
	\includegraphics[scale=0.15]{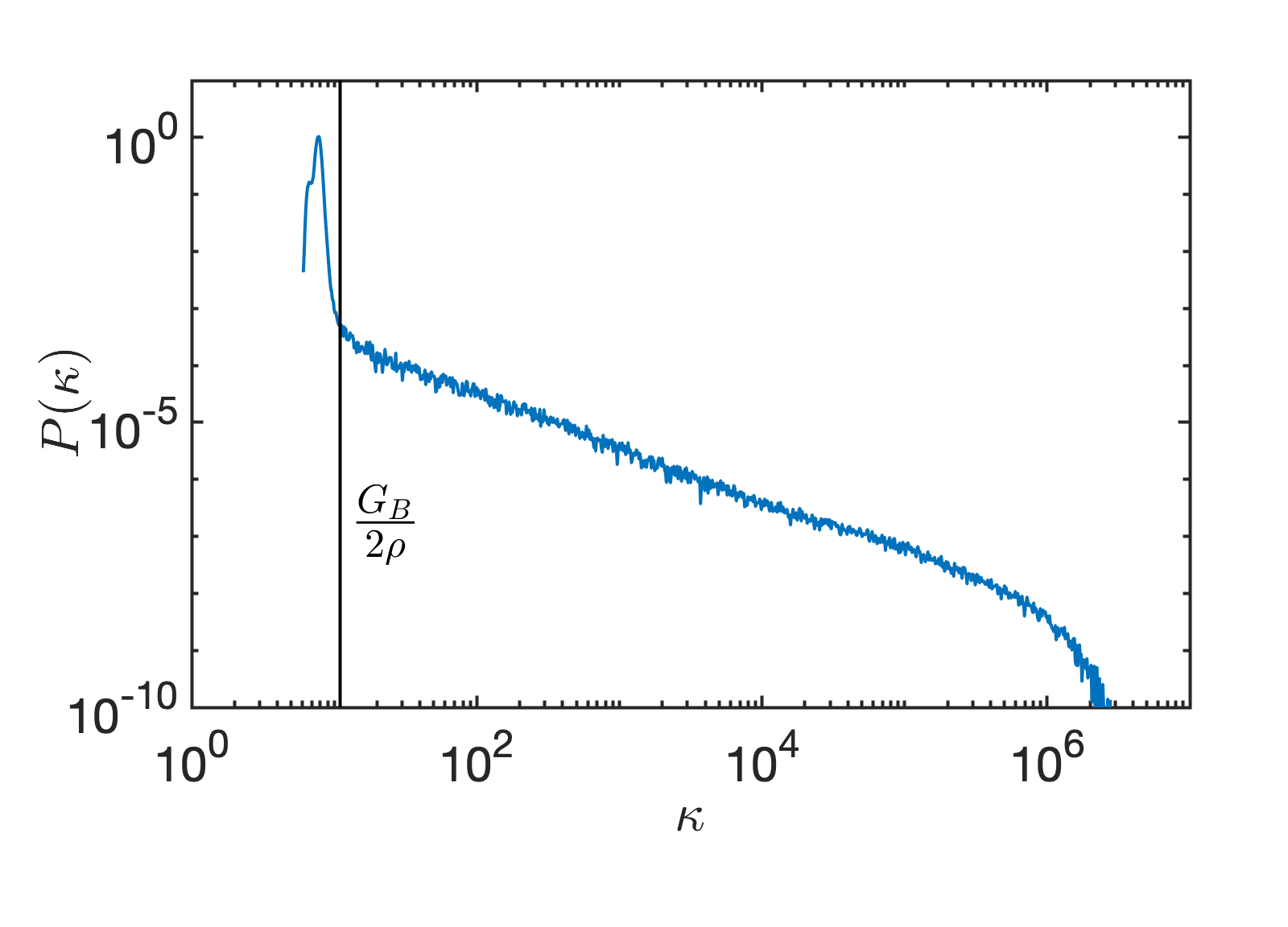}	
\end{center}
\caption{Probability distribution function of $\kappa$ where $\kappa$ has been computed for each strain increment $\delta \gamma$. The vertical solid line represents the lower bound of $\kappa$ to detect plastic events.}
\label{pdf_kappa}
\end{figure}

We determine that $G_B \approx 26$, meaning that the lower bound for plastic activity is $\kappa \approx 13$. To verify our analytical description, we computed $\kappa$ for each strain increment $\delta \gamma=5 \cdot 10^{-5}$ upon deformation (up to $30 \%$). Its probability distribution function is shown in Figure \ref{pdf_kappa}, where we observe that the distribution is peaked around $\kappa \approx 8$ and a long tail persists for larger value of $\kappa$. The peak is associated with elastic contributions. Indeed we measured that $G \approx 12$ for this system so $\kappa$ is more likely to be $(26-12)/2 \approx 7$.

\begin{figure}[t]
\begin{center}
	\includegraphics[scale=0.25]{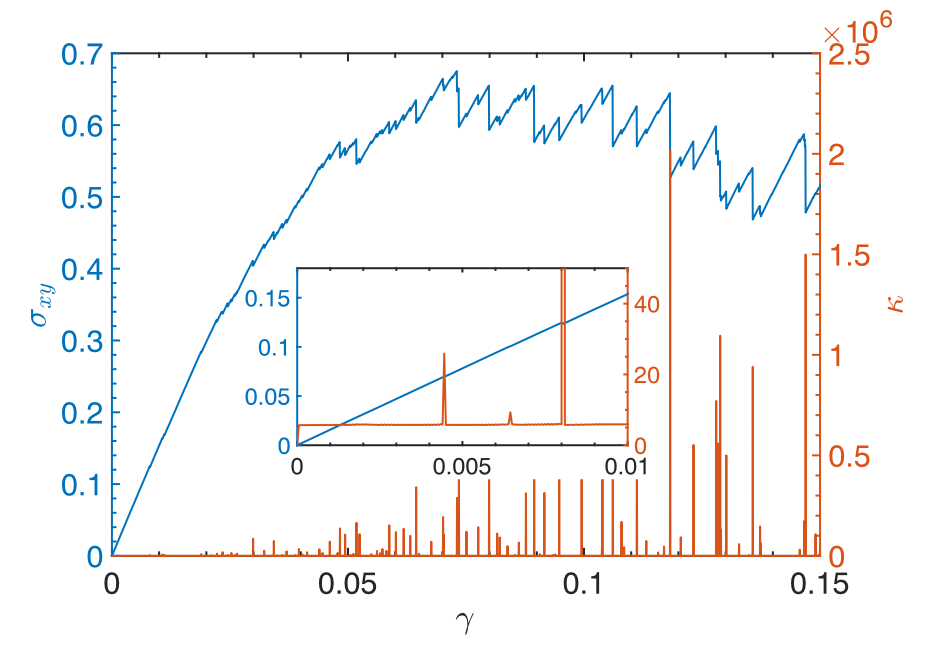}
\end{center}
\caption{Main panel: Typical stress-strain curve obtained during the deformation of a global system where $L=100$. The peaks in the $\kappa$ observable indicate stress drops. Inset: Zoom in the initial stage of the deformation.}
\label{stress_strain_kappa}
\end{figure}

We decided to chose $\kappa \approx 30$ as criterion to ensure to have only plastic contributions. In Figure \ref{stress_strain_kappa}, we observe that $\kappa$ is matching with stress release in the stationary regime but more interestingly, as suggested in the inset, this criterion allows one to determine plastic events for which no stress drop is measured. This is due to the higher sensitivity of the potential energy to structural rearrangement. The detection in terms of stress drops is constrained by the choice of the strain increment as a stress release would have been certainly noticed with a smaller $\delta \gamma$ but this would also imply longer simulation times. Therefore, $\kappa$ is an efficient criterion to monitor plasticity while working with a finite value of $\delta \gamma$. 

\bibliography{references.bib}

\begin{thebibliography}{10}
\bibitem{info_noise}
The statistics for the noise obtained for the smallest system $L=53$ is not
  good enough to extract any reliable information. Indeed as reducing the size
  implies larger $\gamma_Y$ between events, we did not record enough stress
  drops on a strain interval of $2\%$.
\end{thebibliography}
\bibliographystyle{unsrt}

\end{document}